\documentclass[aip,jcp,11pt,reprint]{revtex4-1}
\usepackage{natbib}
\usepackage{amsmath}
\usepackage{booktabs}
\usepackage{graphicx}
\usepackage{color}%
\newcommand{\rv}[1]{\ensuremath{\mathbf{#1}}} 
\newcommand{\gv}[1]{\ensuremath{\mbox{\boldmath$ #1 $}}} 

\begin{document}
\date{\today}

\title{A time-dependent formulation of coupled cluster theory for many-fermion systems at finite temperature}

\author{Alec F. White}
\email{whiteaf@berkeley.edu}
\author{Garnet Kin-Lic Chan}
\email{gkc1000@gmail.com}
\affiliation{Division of Chemistry and Chemical Engineering, California Institute of Technology, Pasadena,
California 91125, USA}

\begin{abstract}
We present a time-dependent formulation of coupled cluster theory. This theory allows for direct computation of the free energy of quantum systems at finite temperature by imaginary time integration and is closely related to the thermal cluster cumulant theory of Mukherjee and co-workers. Our derivation highlights the connection to perturbation theory and zero-temperature coupled cluster theory. We show explicitly how the finite-temperature coupled cluster singles and doubles amplitude equations can be derived in analogy with the zero-temperature theory and how response properties can be efficiently computed using a variational Lagrangian. We discuss the implementation for realistic systems and showcase the potential utility of the method with calculations of the exchange correlation energy of the uniform electron gas at warm dense matter conditions. 
\end{abstract}
\maketitle

\section{Introduction}
In calculations of the electronic structure of molecules and materials, the effects of a finite electronic temperature are usually not considered. This is sufficient for nearly all molecular systems and for many systems in the condensed phase, because only a small number of electronic states are thermally populated at typical temperatures.
However, there are cases where the electronic temperature plays a crucial role.
In correlated electron materials, interactions lead to low-energy electronic excitations and electronic phase transitions\cite{Jarrell2001,Macridin2005,Maier2005,Yang2011,Gull2013}. Electronic free energy differences can also drive structural transitions, both in molecules, such as in spin cross-over complexes\cite{bousseksou2011molecular}, as well as in crystals\cite{Moroni1996}. Hot electrons can be used to drive new kinds of reactions, as seen in hot electron-driven chemistry on plasmonic nano-particles\cite{Mukherjee2013}. And finally, the properties of materials under
extreme conditions\cite{Fortov2009}, including at high electronic temperatures\cite{Ernstorfer2009},  is also of interest for a variety of applications. For all these problems, a quantum many-body theory at finite temperature is required, and this has lead to renewed interest in computational approaches.


The simplest treatment of many-body systems is mean field theory, and mean field theory at finite temperature, in the form of Hartree-Fock\cite{Mermin1963} or density functional theory (DFT)\cite{Mermin1965,Stoitsov1988}, is routinely used. In recent years, experimental interest in matter at high temperatures has spurred much activity in finite temperature DFT.\cite{Eschrig2010,Pittalis2011,Cytter2018,Karasiev2018} However, a description of electron correlations beyond the mean-field/DFT level is often required for accurate computation of chemical and material properties. Methods for the approximate treatment of correlations based on finite-temperature perturbation theory and finite-temperature (Matsubara) Green's functions have been known for many years\cite{Fetter2003,Blaizot1985,Negele1987}, and there has been some recent interest in applying these techniques in an {\it ab initio} context.\cite{Kananenka2016,Welden2016} They are commonly used as impurity solvers within
dynamical mean field theory (DMFT)\cite{Georges1996,Kotliar2006,Gull2013} and the related dynamical cluster approximation (DCA)\cite{Hettler1998,Jarrell2001,Maier2005,Macridin2005,Yang2011}. Finite temperature quantum Monte Carlo (QMC) methods such as determinantal QMC and path integral Monte Carlo (PIMC) have also been studied  for many years\cite{blankenbecler1981monte,Liu2018,Chandler1981,Ceperley1995,Militzer2000}. However, for fermionic systems, QMC methods display a sign problem, limiting simulations to high temperatures, or requiring the introduction of additional constraints, such as the fixed node approximation in PIMC (called restricted PIMC (RPIMC))\cite{Ceperley1991,Brown2013a}. There has been recent work to
explore formulations of QMC where the sign problem is less severe under the conditions of interest,  including the configuration path integral Monte Carlo (CPIMC)\cite{Schoof2011} and density matrix quantum Monte Carlo (DQMC)\cite{Blunt2014,Malone2015}. Much of this research has been motivated by calculations on the uniform electron gas for the benchmarking and/or parameterization of finite temperature density functionals\cite{Karasiev2014,Dornheim2018,Karasiev2018}. We will return to this topic in Section~\ref{sec:ueg}.

The coupled cluster method, widely used for its accuracy at zero temperature\cite{Paldus1972,Bartlett1981,Farnell2004,Bartlett2007,Booth2013}, has not seen widespread application at finite temperatures. Kaulfuss and Altenbokem were the first to try to extend coupled cluster theory to finite temperatures by means of an exponential ansatz for the density matrix\cite{Kaulfuss1986}. However, their formalism requires knowledge of the spectrum of the interacting Hamiltonian and is therefore ill-suited to computations on realistic systems. Mukherjee and coworkers have developed a more practical method which they have termed the thermal cluster cumulant (TCC) method\cite{Sanyal1992,Sanyal1993,Mandal2001,Mandal2002,Mandal2003}. This method is based on a thermally normal ordered exponential ansatz for the interaction picture imaginary-time propagator. The TCC method has a formal similarity to single reference and multi-reference coupled cluster theories, but the applications have been limited to very small systems and semi-analytical problems. Hermes and Hirata have recently presented a finite-temperature coupled cluster doubles (CCD) method\cite{Hermes2015} based on ``renormalized" finite-temperature perturbation theory\cite{Hirata2013}. Hummel has independently developed a time-dependent coupled cluster theory\cite{Hummel2018} which is closely related to Hirata's renormalized perturbation theory. We will discuss some aspects of these methods in Section~\ref{sec:pt}.

In this paper we present an explicitly time-dependent formulation of coupled cluster theory applicable to calculations at zero or finite temperature. Imaginary time integration generates a coupled cluster approximation to the thermodynamic potential in the grand canonical ensemble. This theory, which we will call finite-temperature coupled cluster (FT-CC), represents the finite temperature analogue of traditional coupled cluster in that it has the same diagrammatic content. We highlight this fact by showing how the theory may be derived directly from many-body perturbation theory. This theory is also equivalent to a particular realization of the TCC method. In addition to the theory, we discuss the implementation including analytic derivatives for response properties. Some benchmark calculations are presented as a means of validating the implementation and evaluating the accuracy of the method. Finally, we present calculations of the exchange-correlation energy of the uniform electron gas (UEG) at conditions in the warm dense matter regime.

\section{Theory}

\subsection{Finite temperature coupled cluster equations}\label{sec:ft_cc_eq}
Before discussing the details of the derivation of the FT-CC equations, it is instructive to state the result and discuss the analogy with the zero-temperature theory. Conventional, zero-temperature, coupled cluster theory is described in detail in a variety of reviews and monographs\cite{Bartlett1981,Helgaker2000,Bartlett2007,Shavitt2009}. We will review the basic aspects of the theory in order to facilitate comparison with the finite temperature theory developed in this paper. Recall that the coupled cluster method can be derived from an exponential wavefunction ansatz
\begin{equation}
	|\Psi_{CC}\rangle = e^T|\Phi_0\rangle,
\end{equation}
where $|\Phi_0\rangle$ is a single determinant reference. 
The $T$-operator is defined in some space of configurations, $\{\Phi_{\mu}\}$, such that
\begin{equation}
	T = \sum_{\mu} t_{\mu}a_{\mu}
\end{equation}
where $t_{\mu}$ is an amplitude and $a_{\mu}$ is an excitation operator such that
\begin{equation}
	a_{\mu}|\Phi_0\rangle = |\Phi_{\mu}\rangle.
\end{equation}
Generally, the $T$-operator is truncated at some finite excitation level. For example, letting $T = T_1 + T_2$ yields the coupled cluster singles and doubles (CCSD) approximation. The coupled cluster energy and amplitudes are then determined from a projected Schrodinger equation:
\begin{equation}
	\langle \Phi_0|e^{-T}He^T|\Phi_0\rangle = E_{HF} + E_{CC}
\end{equation}
\begin{equation}
	\langle \Phi_{\mu}|e^{-T}He^T|\Phi_0\rangle = 0.
\end{equation}
These equations can be written explicitly in terms of the $T$-amplitude and molecular integral tensors using diagrammatic methods\cite{Paldus1972,Shavitt2009} or computer algebra\cite{Janssen1991,Hirata2003}. The correlation contribution to the energy has a particularly simple form in terms of the $T_1$ and $T_2$ amplitudes:
\begin{equation}\label{eqn:ccE}
	E_{CC} = \sum_{ia}t_i^a f_{ia} + 
	\frac{1}{4}\sum_{ijab} \langle ij||ab\rangle (t_{ij}^{ab} +2t_i^at_j^b).
\end{equation}
Though this wavefunction-based derivation is usually favored, the resulting energy has a well-understood connection to perturbation theory (See for example Chapters 9.4 and 10.4 of Ref.~\citenum{Shavitt2009}). 

In finite-temperature coupled cluster theory, we use an explicitly time-dependent formulation. The time dependent analogues of the $T$-amplitudes are functions of an imaginary time, $\tau$, and will be denoted by $s_{\mu}(\tau)$. At finite temperature and chemical potential, we denote the coupled cluster contribution to the grand potential as $\Omega_{CC}$ such that, given a particular reference,
\begin{equation}
	\Omega = \Omega^{(0)} + \Omega^{(1)} + \Omega_{CC}.
\end{equation}
The coupled cluster contribution is given by
\begin{eqnarray}\label{eqn:FT_CC_Omega}
	\Omega_{CC} &=& \frac{1}{4\beta}\sum_{ijab}\langle ij||ab\rangle
	\int_0^{\beta}d\tau
	[s_{ij}^{ab}(\tau) + 2s_i^a(\tau)s_j^b(\tau)] \nonumber \\
	&+& \frac{1}{\beta} \sum_{ia}f_{ia}\int_0^{\beta}d\tau s_i^a(\tau)
\end{eqnarray}
with $\beta$ the inverse temperature. In the limit $\beta \rightarrow \infty$ Equation~\ref{eqn:FT_CC_Omega} reduces to
\begin{eqnarray}\label{eqn:FT_CC_Omegalim}
	\lim_{\beta \rightarrow \infty}  \Omega_{CC} &=& \frac{1}{4}
    \sum_{ijab}\langle ij||ab\rangle\lim_{\tau\rightarrow \infty}
	[s_{ij}^{ab}(\tau) + 2s_i^a(\tau)s_j^b(\tau)] \nonumber \\
	&+& \sum_{ia}f_{ia}\lim_{\tau\rightarrow \infty} s_i^a(\tau).
\end{eqnarray}
In this limit, $\Omega \rightarrow E - \mu N$. For an insulator, the correlation contribution to $N$ will vanish at zero temperature assuming that $\mu$ can be chosen such that the non-interacting and correlated system have the same number of particles. This requires the non-interacting and correlated energy gaps to have non-vanishing overlap which is typically the case, from which it follows that
\begin{equation}
	\lim_{\beta \rightarrow \infty} \Omega_{CC} = E_{CC}.
\end{equation}
Comparing Equation~\ref{eqn:FT_CC_Omegalim} with Equation~\ref{eqn:ccE}, it is clear that
\begin{equation}
	\lim_{\tau \rightarrow \infty} s_i^a(\tau) = t_i^a \qquad 
	\lim_{\tau \rightarrow \infty} s_{ij}^{ab}(\tau) = t_{ij}^{ab}.
\end{equation}
This is true as long as both amplitudes correspond to the same solution of the non-linear amplitude equations. This correspondence also implies that the $\beta\rightarrow \infty$ limit of these time-dependent amplitudes is related to the imaginary-time version of the amplitudes that appear in time-dependent, wavefunction-based formulations of coupled cluster\cite{Schonhammer1978,Hoodbhoy1978,Huber2011,Kvaal2012,Nascimento2016}.

The FT-CC amplitude equations closely resemble the amplitude equations of zero temperature coupled cluster, and they are diagrammatically identical as we will discuss in Section~\ref{sec:ft_ccpt}. This allows the equations to be written down in precise analogy with the zero-temperature amplitude equations:
\begin{itemize}
\item replace $t_{\mu}$ with $s_{\mu}(\tau')$
\item for each contraction, sum over all orbitals instead of just occupied or virtual orbitals
\item include an occupation number from the Fermi-Dirac distribution ($n_i$ or $1 - n_a$) with each index not associated with an amplitude
\item multiply each term by $-1$
\item for each term contributing to $s_{\mu}(\tau)$, multiply by an exponential factor $\exp[\Delta_{\mu}(\tau' - \tau)]$ and integrate $\tau'$ from $0$ to $\tau$.
\end{itemize} 
As an example we compare the zero-temperature and finite-temperature versions of a term linear in $T_1$ (or $S_1(\tau')$ at finite temperature) which contributes to $T_2$ (or $S_2(\tau)$ at finite temperature):
\begin{equation}
	t_{ij}^{ab} \leftarrow \frac{1}{\Delta_{ij}^{ab}} 
	P(ij)\sum_{c}\langle ab||cj\rangle t_i^c
\end{equation}
\begin{eqnarray}
	s_{ij}^{ab}(\tau)&\leftarrow& -P(ij)\sum_c(1 - n_a)(1 - n_b)n_j
	\langle ab||cj\rangle  \nonumber \\
	&&\times \int_0^{\tau} d\tau'e^{(\varepsilon_a + \varepsilon_b
	-\varepsilon_i - \varepsilon_j)(\tau' - \tau)}s_i^c(\tau')
\end{eqnarray}
The full FT-CCSD amplitude equations are given in Appendix~\ref{sec:Accsd}. We discuss the origin of these specific rules in Sections~\ref{sec:ft_ccpt} and~\ref{sec:ft_tcc}.

\subsection{Perturbation theory at zero and finite temperature}\label{sec:pt}
Perturbation theory for the many-body problem has a long history in chemistry and physics. Time-independent Rayleigh-Schrodinger perturbation theory, time-dependent (or frequency-dependent) many-body perturbation theory at zero temperature, and imaginary time-dependent (or imaginary frequency-dependent)  many-body perturbation theory at finite temperature are discussed in a variety of monographs\cite{Szabo1982,Blaizot1985,Negele1987,Gross1991,Fetter2003,Shavitt2009}.
For completeness, in Appendix~\ref{sec:Apt} we give explicit rules for the diagrammatic derivation of time-domain expressions for the shift in the grand potential in the form most relevant to coupled cluster theory.
As an example, applying these rules at second order yields
\begin{widetext}
\begin{eqnarray}\label{eqn:Omega2}
	\Omega^{(2)} &=& \frac{1}{4\beta}\sum_{ijab}|\langle ij|| ab\rangle|^2n_in_j
	(1 - n_a)(1 - n_b)\left[
	\frac{\beta}{\varepsilon_i + \varepsilon_j - \varepsilon_a - \varepsilon_b} + 
	\frac{1 - e^{\beta(
	\varepsilon_i + \varepsilon_j - \varepsilon_a - \varepsilon_b)}}
	{(\varepsilon_i + \varepsilon_j - \varepsilon_a - \varepsilon_b)^2}\right]
	\nonumber \\
	&+&\frac{1}{\beta}\sum_{ia}|f_{ai}|^2n_i(1 - n_a)\left[
	\frac{\beta}{\varepsilon_i - \varepsilon_a} +
	\frac{1 - e^{\beta(\varepsilon_i - \varepsilon_a)}}
	{(\varepsilon_i - \varepsilon_a)^2}\right].
\end{eqnarray}
\end{widetext}
In this expression, all sums run over all orbital indices. We use $f_{pq}$ and $\langle pq||rs\rangle$ to indicate the one-particle and anti-symmetrized, two-particle elements of the interaction. We have analytically performed the time integrals to obtain the final, time-independent expressions. 

The terms containing exponential factors vanish when summed. However, one must be careful when evaluating the terms where the energy denominators appear to vanish. Such cases were called ``anomalous" by Kohn and Luttinger\cite{Kohn1960} and they require special consideration to obtain the proper finite result. Since each term is an integral of a non-singular function over a finite interval, each term in the sum should be individually finite. We explicitly include the exponential factors in this discussion so that Equation~\ref{eqn:Omega2} is finite term-by-term for finite $\beta$. The second order correction can diverge as $\beta\rightarrow \infty$, but such divergences are well-known in systems that are metallic at 0th order. In such cases, finite temperature perturbation theory will not reduce to perturbation theory at zero temperature, as first observed by Kohn and Luttinger\cite{Kohn1960}. This is hardly surprising since the two perturbation theories compute different quantities. This is particularly clear if we express the 2nd order energy corrections in terms of derivatives of the exact energy, $E$, with respect to a coupling constant, $\lambda$: 
\begin{equation}
	E_{\mathrm{MP2}} = \left.\frac{\partial^2 E}{\partial\lambda^2}
	\right|_{\lambda = 0, N}\qquad 
	E_{\mathrm{FT-MP2}} = \left.\frac{\partial^2 E}{\partial\lambda^2}
	\right|_{\lambda = 0, \mu}. \label{eq:kl}
\end{equation}
For a metallic system, the derivative at fixed $\mu$ will differ from the derivative at fixed $N$ even as $T\rightarrow 0$,  simply because the chemical potentials of the Hartree-Fock reference system and the interacting system are different. Santra and Schirmer published a pedagogical discussion which elaborates on this particular aspect of finite temperature perturbation theory\cite{Santra2016}. 

In light of this discussion, it is clear that the distinction between the two quantities in Equation~\ref{eq:kl}, termed the Kohn-Luttinger conundrum by Hirata and He\cite{Hirata2013}, does {\it not} imply any particular problem with FT-MBPT; it simply reflects the different conditions under which the partial derivative is taken, from the different ensembles  in the zero- and finite-temperature theories. For this reason, we do not discuss the ``renormalized'' finite-temperature MBPT of Hirata and He\cite{Hirata2013} and the related coupled cluster doubles method\cite{Hermes2015}, which incorrectly modify finite temperature perturbation theory to force these two derivatives to be the same in the limit of zero temperature.


\subsection{Time-dependent coupled cluster from perturbation theory}\label{sec:ft_ccpt}
\begin{figure}
\center
\includegraphics[scale=0.45]{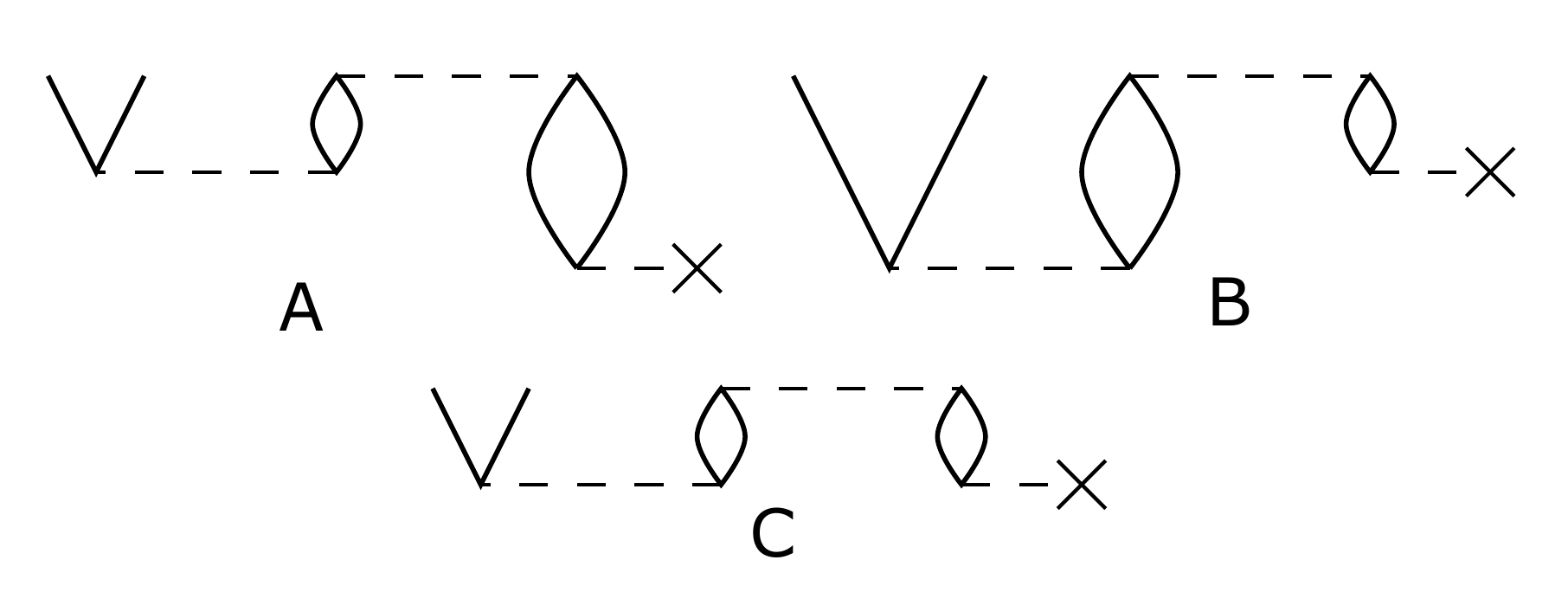}
\caption{\label{fig:diagrams} Different time-orderings of a term relevant to CCSD. }
\end{figure}

The interpretation of coupled cluster theory in the context of many-body perturbation theory can be used to directly define FT-CC theory. The essential point is to require that the energy and amplitude equations reproduce exactly the diagrammatic content of the zero-temperature theory. However, the time-dependent perturbation theory will in general necessitate the consideration of different time orderings. Consider the open diagrams shown in Figure~\ref{fig:diagrams} as an example. 
We must consider both diagrams A and B, and each corresponds to nested integrals of the form  
\begin{eqnarray}
	A &\sim& v_{bc}^{jk}(\tau)\int_0^{\tau}d\tau' f_j^b(\tau') \int_0^{\tau'}
	d\tau''v_{ki}^{ca}(\tau'') \\
	B &\sim& v_{bc}^{jk}(\tau)\int_0^{\tau}d\tau'v_{ki}^{ca}(\tau')\int_0^{\tau'}
	d\tau''f_j^b(\tau'')
\end{eqnarray}
where we have omitted the summation and the factors of occupation numbers which will be common in both terms. We have used $v^{pq}_{rs}(\tau)$ and $f_{pq}(\tau)$ to represent the one and two-electron matrix elements in the interaction picture:
\begin{eqnarray}
	v^{pq}_{rs}(\tau) &\equiv&
	\langle pq||rs\rangle e^{(\varepsilon_p+\varepsilon_q 
	- \varepsilon_r - \varepsilon_s)\tau} \nonumber \\
	f_{pq}(\tau) &\equiv&  f_{pq}e^{(\varepsilon_p - \varepsilon_q)\tau}.
\end{eqnarray} 
These nested integrals can be simplified in a manner analogous to the factorization of perturbation theory denominators in coupled cluster at zero temperature (See Chapters 5-6 of Ref.~\citenum{Shavitt2009}). By defining 
\begin{equation}
	V^{pq}_{rs}(\tau) \equiv \int_0^{\tau} d\tau'v_{rs}^{pq}(\tau') \qquad 
	F_{pq}(\tau) \equiv  \int_0^{\tau} d\tau' f_{pq}(\tau'),
\end{equation}
such that
\begin{equation}
	f_{pq}(\tau) = \frac{d}{d\tau}F_{pq}(\tau) \qquad 
	v_{rs}^{pq}(\tau) = \frac{d}{d\tau}V^{pq}_{rs}(\tau),
\end{equation}
the reverse of the product rule can be applied to the sum of the two time orderings to yield an expression where all quantities are evaluated at a single time
\begin{equation}
	A + B \propto v_{bc}^{jk}(\tau) F_j^b(\tau)V_{ki}^{ca}(\tau)
\end{equation}
which we represent as diagram $C$ of Figure~\ref{fig:diagrams}. This is the time-domain equivalent of the denominator factorization that allows zero-temperature coupled cluster diagrams to be written without regard to the ordering of the different factors of $T$. Given this factorization, we may define $S$-amplitudes at first order such that
\begin{eqnarray}
	s_i^a(\tau)^{[1]}_I &\equiv& -n_i(1-n_a)F_{ai}(\tau) \\
	s_{ij}^{ab}(\tau)^{[1]}_I &\equiv& - n_in_j(1 - n_a)(1 - n_b)V_{ij}^{ab}(\tau)
\end{eqnarray}
where we use the subscript $I$ to emphasize that we are using the interaction picture. The superscript indicates that they are first order in the interaction. The finite temperature coupled cluster equations at some truncated order (usually singles and doubles) then follow directly from their diagrammatic representation. This guarantees by construction that the FT-CC amplitude equations reproduce exactly the diagrammatic content of the corresponding zero temperature theory.

For the purposes of this derivation, we have used the interaction picture. However, there is a numerical difficulty associated with the time-dependent exponential factors which, at long times, will be become exponentially large or small. This leads to problems of overflow or underflow when storing the amplitudes as floating point numbers. This difficulty can be largely overcome by moving to the Schrodinger picture:
\begin{equation}
	s_{\mu}(\tau) \equiv s_{\mu}(\tau)_Ie^{-\Delta_{\mu}\tau}.
\end{equation}
At first order, the Schrodinger-picture singles and doubles amplitudes are proportional to the Schrodinger-picture matrix elements which are time-independent in the usual case. Furthermore, these amplitudes are well-behaved in the limit as $\tau \rightarrow \infty$ in that they reduce to the zero temperature coupled cluster amplitudes. The FT-CCSD amplitude equations for the Schrodinger-picture amplitudes are given in Appendix~\ref{sec:Accsd}.

\subsection{Relationship to thermal cluster cumulant theory}\label{sec:ft_tcc}
The finite temperature coupled cluster method that we have presented here can also be viewed as a particular realization of the thermal cluster cumulant (TCC) theory developed by Mukherjee and others\cite{Sanyal1992,Sanyal1993,Mandal2001,Mandal2002,Mandal2003}. If we denote the thermal normal ordering of a string of operators by $N[\ldots]_0$, then the TCC method uses a normal-ordered ansatz for the imaginary-time propagator: 
\begin{equation}\label{eqn:tcc_ansatz}
	U_I(\tau) = N\left[e^{S(\tau) + X(\tau)}\right]_0.
\end{equation}
Here, $S(\tau)$ is an operator and $X(\tau)$ is a number. The imaginary time propagator obeys a Bloch equation,
\begin{equation}\label{eqn:Bloch}
	-\frac{\partial U_I}{\partial \tau } = V_I(\tau)U_I(\tau),
\end{equation}
from which differential equations for $S(\tau)$ and $X(\tau)$ may be determined. The expression for the thermodynamic potential follows directly from the ansatz of Equation~\ref{eqn:tcc_ansatz}:
\begin{equation}\label{eqn:tccE1}
	\Omega = \Omega^{(0)} - \frac{1}{\beta}X(\beta).
\end{equation}

As shown in Ref.~\citenum{Sanyal1993}, Equation~\ref{eqn:Bloch} implies coupled differential equations for $S$ and $X$. Solving these equations by integration yields the FT-CC equations
\begin{equation}\label{eqn:tccE2}
	X(\tau) = -\tau\Omega^{(1)} - \int_0^{\tau}d\tau' 
    \left[V_I^N(\tau)N[e^{S(\tau)}]_0
	\right]_{\mathrm{fully-contracted}} 
\end{equation}
\begin{equation}\label{eqn:tccS}
	S(\tau) = - \int_0^{\tau}d\tau' \left[V_I^NN[e^{S(\tau)}]_0\right]_C.
\end{equation}
$V_I^N(\tau)$ is the thermally normal-ordered component of the interaction, and the first order contribution to the free energy, $\Omega^{(1)}$, is the number component of $V$. The subscript $C$ in Equation~\ref{eqn:tccS} indicates that we only consider terms in which $V$ is connected to all the amplitudes by at least one contraction. Inserting Equation~\ref{eqn:tccE2} into Equation~\ref{eqn:tccE1} yields the first order contribution to the grand potential plus the interaction picture version of the FT-CC contribution to the grand potential (Equation~\ref{eqn:FT_CC_Omega}). A minor difference is that in our formulation we have absorbed the occupation numbers into the definition of the $S$-amplitudes, whereas in the TCC method the occupation numbers arise as a result of thermal contractions involving the $S$ operators. The connected cluster form of Equation~\ref{eqn:tccS} leads to the same set of diagrams obtained in coupled cluster. When properly interpreted, these diagrams reproduce the FT-CC amplitude equations in the interaction picture. Using
\begin{equation}
	S(\tau) = S_1(\tau) + S_2(\tau)
\end{equation}
leads to the FT-CCSD method we have described. 

\subsection{Response properties}
The primary utility of the thermodynamic potential is that differentiation will generate ensemble averages. In practice we most often require the average energy, entropy, and number of particles:
\begin{equation}
	\langle E \rangle = \Omega + T\langle S\rangle + \mu \langle N \rangle
\end{equation}
\begin{equation}\label{eqn:thermprop}
	\langle S\rangle = -\frac{\partial \Omega}{\partial T}  \qquad
	\langle N \rangle = -\frac{\partial \Omega}{\partial \mu}.
\end{equation}
The partial derivatives in Equation~\ref{eqn:thermprop} are partial thermodynamic derivatives but still require the inclusion of the response of any parameters which determine the form of $\Omega$. In general, an observable corresponding to an operator $O$ can be computed by defining a new Hamiltonian
\begin{equation}
	H[\alpha] \equiv H + \alpha O
\end{equation}
and taking the derivative of the thermodynamic potential
\begin{equation}
	\langle O \rangle = \left.\frac{d\Omega[\alpha]}{d\alpha}\right|_{\alpha = 0}.
\end{equation}

Just like the coupled cluster energy at zero temperature, $\Omega_{CC}$ is not a variational function of the amplitudes. This complicates the implementation of analytic derivatives, but this difficulty can be largely mitigated by using a variational Lagrangian as in the zero temperature theory\cite{Salter1989,Helgaker2000,Shavitt2009}. The finite temperature free-energy and amplitude equations have the form
\begin{eqnarray}
	s_{\mu}(\tau) + \int_0^{\tau}d\tau' e^{\Delta_{\mu}(\tau' - \tau)}
    \mathrm{S}_{\mu}(\tau')	&=& 0 \\
	\frac{1}{\beta}\int_0^{\beta} \mathrm{E}(\tau) &=& \Omega_{CC}.
\end{eqnarray}
The precise forms of E and S are given in Appendix~\ref{sec:Accsd}. The computation of properties can be simplified by defining a Lagrangian, $\mathcal{L}$, with Lagrange multipliers $\lambda^{\mu}(\tau)$
\begin{align}
	\mathcal{L} &\equiv \frac{1}{\beta}\int_0^{\beta}\mathrm{E}(\tau) \nonumber \\
	&-
	\frac{1}{\beta}\int_0^{\beta} d\tau \lambda^{\mu}(\tau)\left[
	s_{\mu}(\tau) + \int_0^{\tau}d\tau' e^{\Delta_{\mu}(\tau' - \tau)} \mathrm{S}_{\mu}(\tau')\right]
\end{align}
such that variational optimization of $\mathcal{L}$ with respect to the $\lambda$-amplitudes yields the FT-CC amplitude equations. Variational optimization with respect to the $S$-amplitudes yields equations for $\lambda^{\mu}$. The solution of the FT-CC $\lambda$-equations is discussed in Appendix~\ref{sec:Alambda}.

Once the $\lambda$-amplitudes have been determined, any first order property may be computed from the partial derivative of the Lagrangian. In practice, the specifics of the numerical evaluation of the time integrals must be considered. Some details of the implementation of analytic derivatives are discussed in Appendix~\ref{sec:Alambda}.  

\section{Implementation}
We have developed a simple pilot implementation of FT-CCSD interfaced to the PySCF electronic structure package\cite{Sun2018}.
In our implementation, the numerical integration is performed on a uniform grid using Simpson's rule for the quadrature weights (see Appendix~\ref{sec:Aint} for details). Though effective at high temperatures, this integration scheme is far from optimal at low temperatures and can be improved considerably by taking into account the structure of the $S$ amplitudes at low temperature. For example, we know that
\begin{eqnarray}
	\lim_{\tau\rightarrow 0} s_{\mu}(\tau) &=& 0 \\
	\lim_{\tau\rightarrow \infty} s_{\mu}(\tau) &=& [\mathrm{const.}],
\end{eqnarray}
and this information can be used to develop much more efficient quadrature schemes at low temperatures. However, we have not pursued this in this work.

In our implementation, the integrals are contracted with the occupation numbers once before the start of the iterations. A guess for the $S$-amplitudes is obtained from the MP2 amplitudes or from a previous calculation. Using the modified integrals and the guess, the coupled cluster iterations proceed in two steps. First, $\text{S1}_i^a(\tau')$ and $\text{S2}_{ij}^{ab}(\tau')$ of Equations~\ref{eqn:S1} and~\ref{eqn:S2} are evaluated at each time point. Second, these quantities are integrated as described in Appendix~\ref{sec:Aint} to obtain new amplitudes. In our implementation we compute the amplitudes for all times at each iteration. It is possible to invert this algorithm so that the amplitudes are converged in a point-by-point manner starting with $\tau = 0$. The number of iterations needed to achieve convergence is strongly temperature dependent: more iterations are generally required at lower temperatures. In practice it is also sometimes necessary to damp the iterations to achieve convergence at lower temperatures. Direct inversion of the iterative subspace (DIIS) convergence acceleration\cite{Pulay1980,Pulay1982,Scuseria1986} could potentially be used to speed up convergence at the cost of additional storage.

We used the formulation of Stanton and Gauss\cite{Stanton1991}  to implement the amplitude equations efficiently. Similar intermediates are used in the solution of the $\lambda$-equations. At low temperatures, the FT-CCSD equations can be somewhat simplified in that summations over all orbitals can be restricted to those terms where the products of occupation numbers are non-negligible. In other words, if $1 - n_i$ or $n_a$ are small enough, some terms can be ignored in the sums. Unfortunately, this threshold must be very tight in practice, and this simplification did not provide any noticeable gains for the systems considered in this study. However, this approximation will be absolutely necessary in the limit as $\beta\rightarrow\infty$ to prevent overflow.

\section{Results}

\subsection{Benchmark calculation}
In order to validate the implementation of the method and test its accuracy, we report calculations on an exactly solvable system: Be atom in a minimal basis. It does not make physical sense to consider a vacuum system in the grand canonical ensemble, but the model is nonetheless well-defined in a finite basis. This model system involves 5 spatial orbitals and thus can be solved exactly. In the grand canonical ensemble an exact solution requires, at least in principle, tracing over all possible particle number and spin sectors. In all calculations we use the orbitals computed at zero temperature.

\begin{figure}
\center
\includegraphics{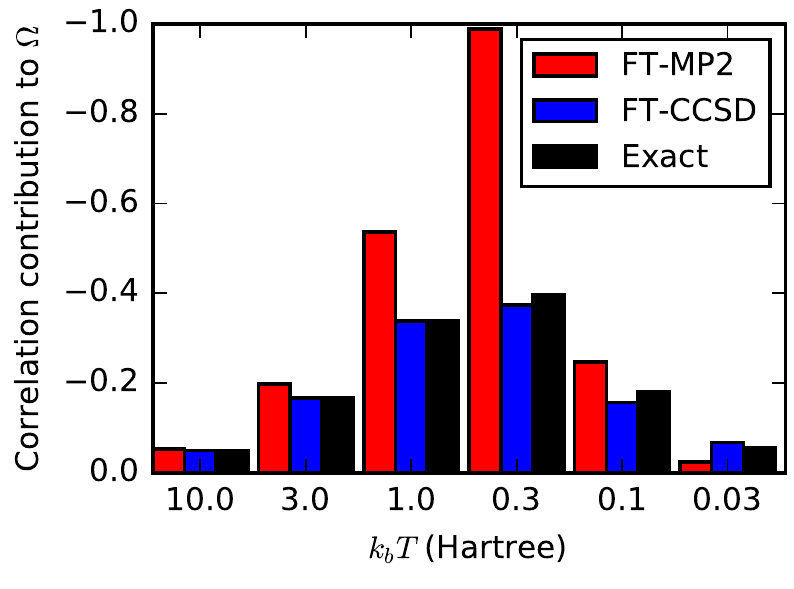}
\caption{\label{fig:Gcompare} The FT-MP2, FT-CCSD, and exact correlation contributions to the grand potential in $E_h$ for the Be model. FT-CCSD at worst underestimates the correlation contribution by $\sim 13\%$. }
\end{figure}
For this particular system, FT-CCSD performs very well. Figure~\ref{fig:Gcompare} shows the correlation contribution to the thermodynamic potential computed with FT-MP2, FT-CCSD, and exact diagonalization. The temperature range was chosen to be high enough that the finite temperature effects are quite significant, but not so high that the non-interacting system becomes exact. FT-CCSD universally outperforms FT-MP2, as we might expect, and the energies are at worst in error by $~13\%$. The good performance of FT-CCSD persists even in the problematic cases where $\Omega^{(2)}$ is a significant overestimate of the exact correlation contribution.

\begin{figure}
\center
\includegraphics{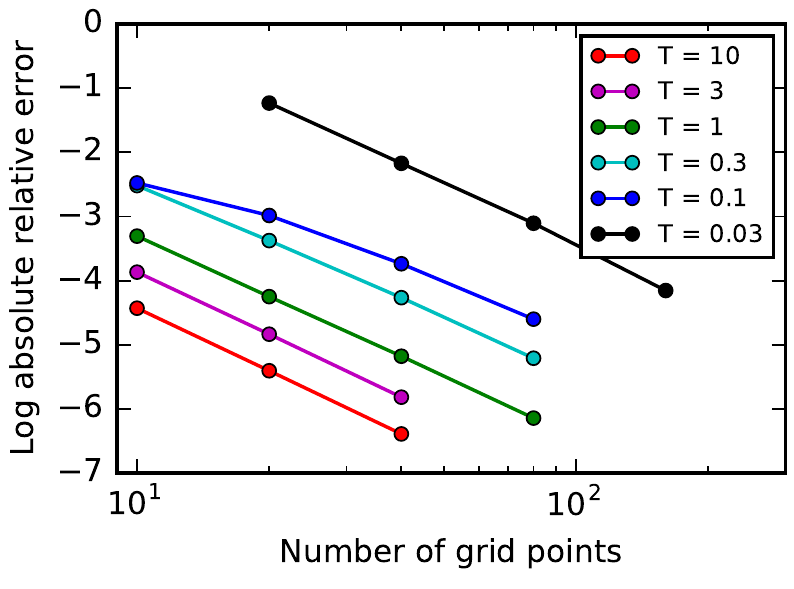}
\caption{\label{fig:grid} The convergence of the correlation contribution to the grand potential of the Be model with respect to the size of integration grid for different temperatures. }
\end{figure}
We have also used this model system to study the convergence with respect to the grid used for numerical integration. The relative error in the computed value $\Omega_{CC}$ due to numerical integration is shown in Figure~\ref{fig:grid} as a function of the number of grid points. The number of grid points required to obtain a specified accuracy depends strongly on the temperature. In general, it will also depend on the energy spectrum of the particular problem. In this case, acceptable accuracy can be obtained at high temperatures ($k_BT \geq 1.0$ $E_{\text{h}}$) with $\sim 10$ grid points. At lower temperatures more grid points are required, and in practice one should ensure convergence of the property of interest with respect to the quadrature grid. Also, we have observed that the amplitude equations require less damping and converge in fewer iterations when more grid points are used.

\subsection{The uniform electron gas at finite temperature}\label{sec:ueg}
The regime of ``warm dense matter" has been the subject of much recent theoretical and experimental interest\cite{Fortov2009,Shukla2011,Dornheim2018}. Warm dense matter is loosely characterized by an electron Wigner-Seitz radius, $r_s$, and reduced temperature, $\theta = k_BT/E_F$, both of order 1. The theoretical description of matter under these conditions is challenging due to the similar importance of thermal effects and quantum exchange and correlation. The uniform electron gas at warm dense matter conditions has emerged as an essential test for theory and an ingredient for the parameterization of various flavors of finite temperature DFT\cite{Gupta1980,Dharma-Wardana1981,Karasiev2014,Karasiev2016,Karasiev2018}.
Ref.~\citenum{Dornheim2018} offers a comprehensive review which highlights progress in quantum Monte Carlo (QMC) calculations in particular. In the past, some calculations have been reported in the grand canonical ensemble\cite{Gupta1980,Dharma-Wardana1981,Perrot1984,Schweng1991}, but recent work has focused on high quality QMC calculations on both the polarized\cite{Brown2013a,Schoof2015a,Groth2016,Malone2016} and unpolarized\cite{Brown2013a,Dornheim2016,Dornheim2016a} UEG in the canonical ensemble. The fixed node approximation of RPIMC is a source of uncontrolled error\cite{Filinov2014}, and since the work of Brown {\it et al}\cite{Brown2013a}, there has been considerable effort to obtain more accurate results over a wider range of $r_s$\cite{Blunt2014,Malone2015,Filinov2015,Schoof2015a,Dornheim2015,Schoof2015,Dornheim2016,Dornheim2016a,Malone2016,Groth2016,Dornheim2016a}. In these studies the $N = 33$ polarized UEG and $N = 66$ unpolarized UEG have emerged as benchmark systems.

\begin{figure}
\center
\includegraphics{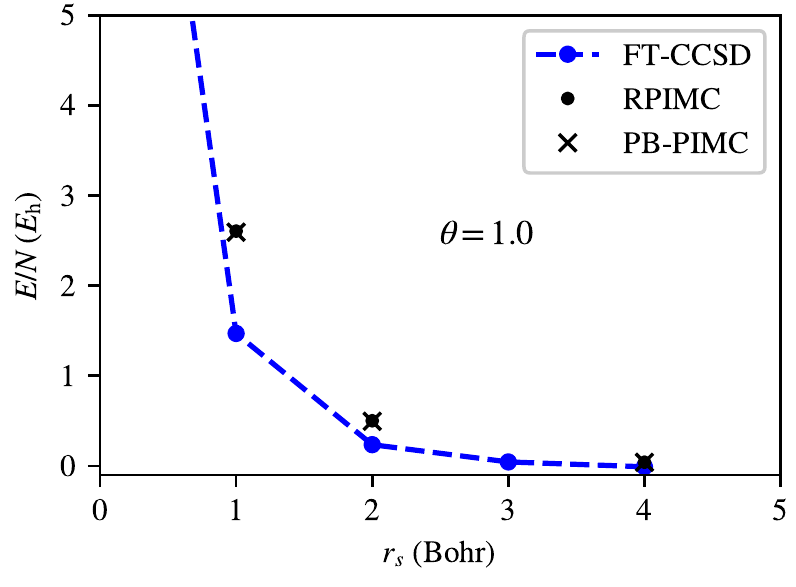}
\includegraphics{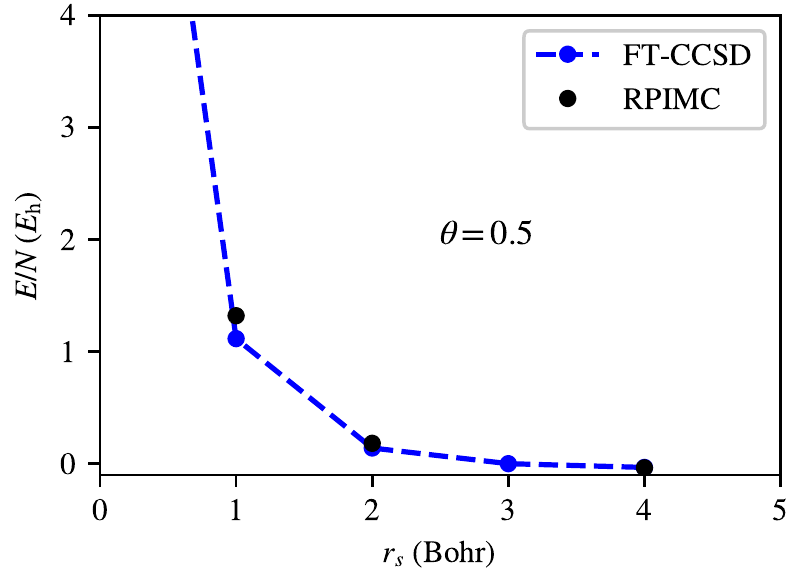}
\includegraphics{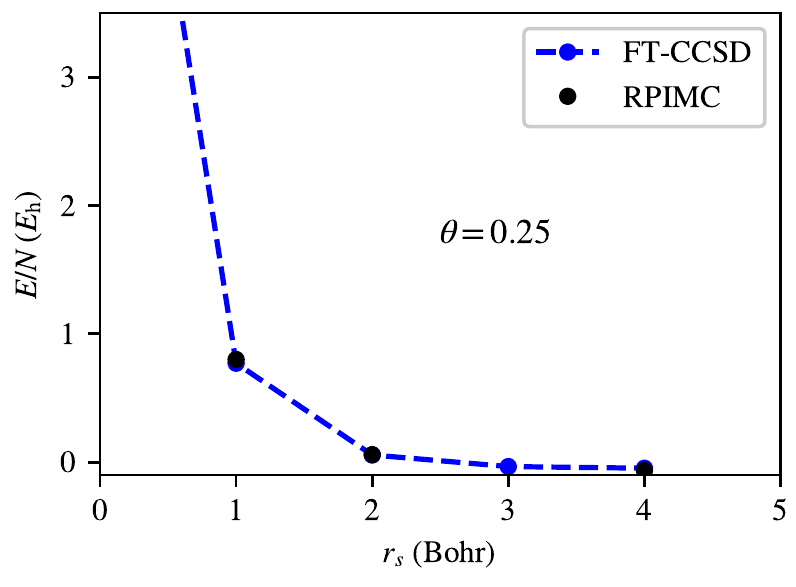}
\caption{\label{fig:ueg} Total energies per electron of the uniform electron gas computed with FT-CCSD. The RPIMC results are those of Brown {\it et al}\cite{Brown2013a}.}
\end{figure}
\begin{figure}
\center
\includegraphics{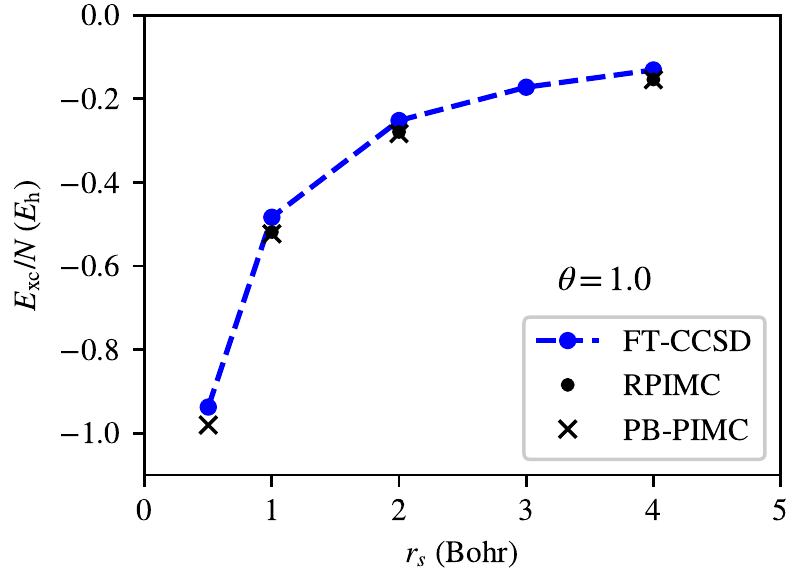}
\includegraphics{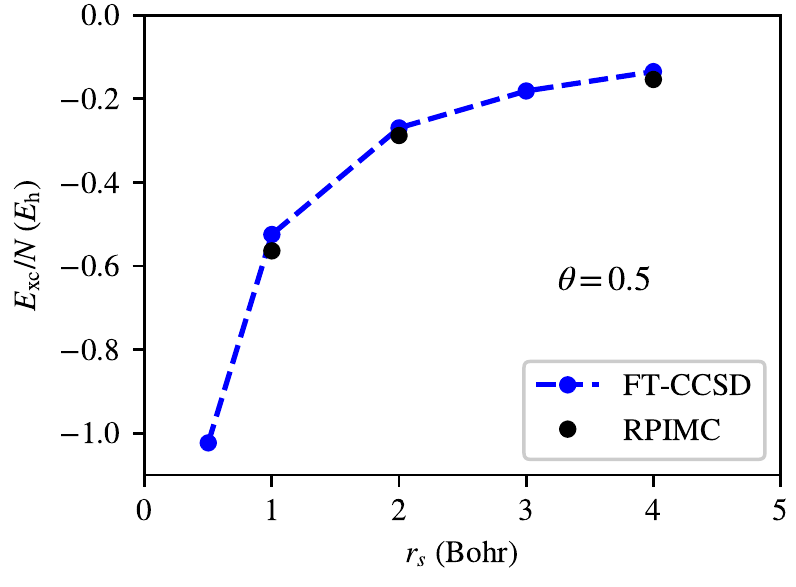}
\includegraphics{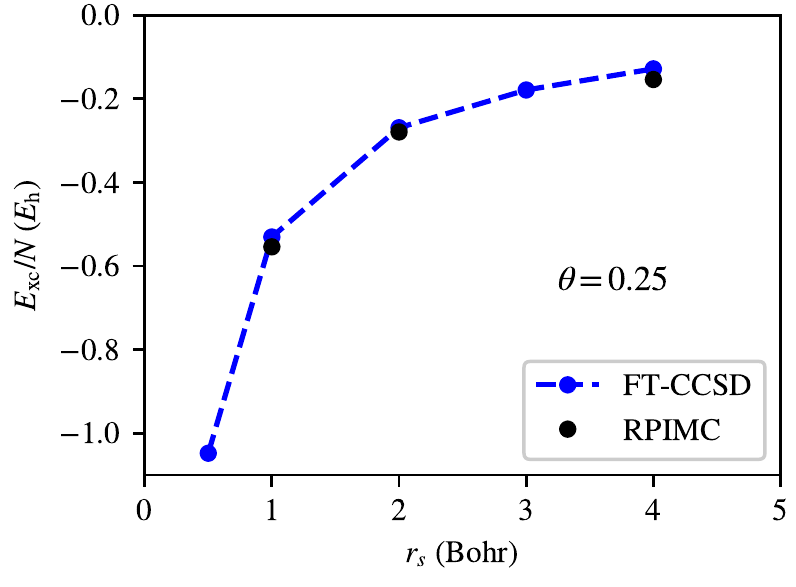}
\caption{\label{fig:ueg_xc} Exchange-correlation energies per electron of the uniform electron gas computed with FT-CCSD. The RPIMC results are those of Brown {\it et al}\cite{Brown2013a}.}
\end{figure}

In Figure~\ref{fig:ueg} we show the total energy per electron of the unpolarized UEG computed with FT-CCSD for several relevant values of $r_s$ and $\theta$. We use a basis of 57 plane waves and the chemical potential is adjusted so that $N=38$. This one-dimensional root finding problem, $N(\mu) - 38 = 0$, is solved with the secant method and takes 4-5 iterations on average. 10 grid points are used for all calculations. The error due to the finite grid will be maximal at low temperatures and at large $r_s$, but even for $r_s = 4$ and $\theta = 0.25$, we estimate the impact of this error on the exchange correlation energy be less than 1\%. Comprehensive tables of all our results are given in the supporting information where we also show results for $N = 14$ and $N = 66$ electrons.
In Figure~\ref{fig:ueg_xc} we show the exchange-correlation energy for the warm-dense UEG. We also offer comparisons with RPIMC calculations\cite{Brown2013a} for all temperatures and permutation-blocking PIMC\cite{Dornheim2016a} for $\theta = 1$. Note that, while the fixed node approximation of RPIMC leads to significant errors for the polarized UEG\cite{Schoof2015a,Dornheim2015}, the fixed node error for the unpolarized UEG is much less severe\cite{Dornheim2016a}. Therefore, RPIMC provides a reasonable benchmark for the range of $r_s$ presented here. 

Note that the QMC and FT-CCSD calculations compute different quantities, as canonical and grand-canonical ensemble results will only agree in the thermodynamic limit, and finite size effects in both cases are large. In addition, the FT-CC works within a (small) orbital basis, while both QMC simulations have no basis set error. Nonetheless, the comparison between the two shows that the equation of state is remarkably similar. Thus we expect FT-CCSD to become a promising tool for the study of warm dense matter.

\section{Conclusions}
In this work we have shown how an explicitly time-dependent formulation of coupled cluster can be used to develop a finite temperature coupled cluster theory. The resulting FT-CC theory can be derived directly from many-body perturbation theory and is formally equivalent to the normal-ordered ansatz of the TCC method. In addition to the derivation of the FT-CCSD amplitude equations, we have also shown how first-order properties may be computed as analytic derivatives using a variational Lagrangian. Preliminary calculations on the uniform electron gas show that FT-CC methods are promising candidates for non-perturbative, non-stochastic computation of the properties of quantum systems at finite temperature.

For large-scale application, a variety of practical improvements are still necessary:
\begin{itemize}
	\item Specialization to restricted reference
	\item Use of disk to lower memory footprint
	\item MPI parallelization over time points
	\item More stable iteration of the amplitude/$\lambda$ equations
\end{itemize}
These improvements mimic the algorithmic advances that have made efficient, black-box implementation of modern coupled cluster methods feasible. There is also further room for improvement in the low temperature regime where the simple structure of the $S$-amplitudes should allow for a reduction of the computational cost.

Finally, it should be noted that the time-dependent formulation of coupled cluster presented here is remarkably general. We have shown how it can be used to unify coupled cluster, thermal cluster cumulant, and many-body perturbation theories into a computational method well-suited to practical implementation. However, further generalizations including the extension to systems out of equilibrium, are possible and are the subject of current investigation.   

\begin{acknowledgements}
The authors would like to thank Jiajun Ren for helpful discussions. This work is supported by the US Department of Energy, Office of Science, via grant number SC0018140.
\end{acknowledgements}

\section*{Supporting information}
Supporting information including all of our raw data on the warm dense UEG is available.

\appendix
\section{FT-CCSD amplitude equations}\label{sec:Accsd}
The FT-CCSD contribution to the thermodynamic potential, given in Equation~\ref{eqn:FT_CC_Omega}, can be written as
\begin{equation}
	\Omega_{CC} = \frac{1}{\beta}\int_0^{\beta} \mathrm{E}(\tau) 
\end{equation}
where
\begin{equation}
	\mathrm{E}(\tau) = \sum_{ia}f_{ia}s_i^a(\tau) + 
    \frac{1}{4}\sum_{ijab}\langle ij||ab\rangle 
    [s_{ij}^{ab}(\tau) + 2s_i^a(\tau)s_j^b(\tau)].
\end{equation}
Note the analogy to the standard, zero-temperature, coupled cluster energy expression. The singles and doubles equations similarly have the simple form
\begin{equation}
	s_i^a(\tau) = -\int_0^{\tau}d\tau'e^{(\varepsilon_a - \varepsilon_i)
	(\tau' - \tau)}\mathrm{S1}_i^a(\tau') 
\end{equation}
\begin{equation}
	s_{ij}^{ab}(\tau) = -\int_0^{\tau}d\tau'e^{(\varepsilon_a +\varepsilon_b
	- \varepsilon_i-\varepsilon_j)(\tau' - \tau)}\mathrm{S2}_{ij}^{ab}(\tau') 
\end{equation}
where the integrands, S1 and S2, are precisely the equations of a zero-temperature CCSD iteration except that each open line that connects to a Hamiltonian fragment carries with it an occupation number:

\begin{widetext}
\begin{eqnarray}
	S1_i^a(\tau') &=& (1 - n_a)n_if_{ai} + \sum_b (1 - n_a)f_{ab}s_i^b(\tau') - 
	\sum_j n_if_{ji}s_j^a(\tau') + 
	\sum_{jb}\langle ja||bi\rangle s_j^b(\tau') \nonumber \\ 
	&+& \sum_{jb}f_{jb}s_{ij}^{ab}(\tau')
	+ \frac{1}{2}\sum_{jbc}(1 - n_a)\langle aj||bc \rangle s_{ij}^{bc}(\tau')
	- \frac{1}{2}\sum_{jkb}n_i\langle jk||ib\rangle s_{jk}^{ab}(\tau')
	-\sum_{jb} f_{jb}s_i^b(\tau')s_j^a(\tau') \nonumber \\
	&+& \sum_{jbc}(1 - n_a)\langle ja||bc \rangle s_j^b(\tau')s_i^c(\tau') 
	- \sum_{jkb}\langle jk||bi\rangle s_j^b(\tau')s_k^a(\tau')
	- \frac{1}{2}\sum_{jkbc}\langle jk||bc\rangle s_i^b(\tau')s_{jk}^{ac}(\tau')
	\nonumber \\ &-& 
	\frac{1}{2}\sum_{jkbc}\langle jk||bc\rangle s_j^a(\tau') s_{ik}^{bc}(\tau') 
	+ \sum_{jkbc}\langle jk||bc\rangle s_j^b(\tau')s_{ki}^{ca}(\tau')
	+ \sum_{jkcd}\langle jk||bc\rangle s_i^b(\tau')s_j^c(\tau')s_k^a(\tau')\label{eqn:S1}
\end{eqnarray}
\begin{eqnarray}
	S2_{ij}^{ab}(\tau') &=& n_in_j(1 - n_a)(1 - n_b)\langle ab||ij\rangle 
	+ P(ij)\sum_{c}n_j(1 - n_a)(1 - n_b)\langle ab||cj\rangle s_i^c(\tau')
	\nonumber \\
	&-& P(ab)\sum_{k}n_in_j(1 - n_b)\langle kb||ij\rangle s_k^a(\tau') + 
	P(ab)\sum_{c}(1 - n_b)f_{bc}s_{ij}^{ac}(\tau') - 
	P(ij)\sum_{k}n_jf_{kj} s_{ik}^{ab}(\tau') \nonumber \\ 
	&+& \frac{1}{2}\sum_{cd}(1 - n_a)(1 - n_b)\langle ab||cd\rangle 
	s_{ij}^{cd}(\tau')
	+\frac{1}{2}\sum_{kl}n_in_j\langle kl||ij\rangle s_{kl}^{ab}(\tau')\nonumber\\
	&+&  P(ij)P(ab)\sum_{kc}n_j(1 - n_b)\langle kb||cj\rangle s_{ik}^{ac}(\tau')
	+\frac{1}{2}P(ij)\sum_{cd}(1 - n_a)(1 - n_b)
	\langle ab||cd\rangle s_i^c(\tau')s_j^d(\tau')\nonumber \\
	&+& \frac{1}{2}P(ab)\sum_{kl}n_in_j\langle kl||ij\rangle 
	s_k^a(\tau')s_l^b(\tau')
	- P(ij)P(ab)\sum_{kc}(1-n_a)n_j\langle ak||cj\rangle s_i^c(\tau')s_k^b(\tau')\nonumber \\
	&-&P(ij)\sum_{kc}f_{kc}s_i^c(\tau')s_{kj}^{ab}(\tau')
	- P(ab)\sum_{kc}f_{kc}s_k^a(\tau')s_{ij}^{cb}(\tau') + 
	P(ab)\sum_{kcd}(1-n_a)\langle ka||cd\rangle s_k^c(\tau')s_{ij}^{db}(\tau') 
	\nonumber \\ &-& 
	P(ij)\sum_{klc}n_i\langle kl||ci\rangle s_k^c(\tau')s_{lj}^{ab}(\tau')
	+ P(ij)P(ab)\sum_{kcd}(1 - n_a)\langle ak||cd\rangle s_i^c(\tau')s_{kj}^{db}
	(\tau') \nonumber \\ 
	&-& P(ij)P(ab)\sum_{klc}n_i\langle kl||ic\rangle s_k^a(\tau')
	s_{lj}^{cb}(\tau') 
	+ \frac{1}{2}P(ij)\sum_{klc} n_j\langle kl||cj\rangle 
	s_i^c(\tau')s_{kl}^{ab}(\tau')\nonumber \\
	&-& \frac{1}{2}P(ab)\sum_{kcd}(1 - n_b)\langle kb||cd\rangle 
	s_k^a(\tau')s_{ij}^{cd}(\tau')
	+ \frac{1}{4}\sum_{klcd} \langle kl||cd\rangle s_{ij}^{cd}(\tau')s_{kl}^{ab}(\tau') \nonumber \\
	&+& \frac{1}{2}P(ij)P(ab)\sum_{klcd}\langle kl||cd\rangle s_{ik}^{ac}(\tau')s_{lj}^{db}(\tau')
	- \frac{1}{2}P(ab)\sum_{klcd} \langle kl||cd\rangle s_{kl}^{ca}(\tau')s_{ij}^{db}(\tau')\nonumber \\
	&-& \frac{1}{2}P(ij)\sum_{klcd}\langle kl||cd\rangle s_{ki}^{cd}(\tau')s_{lj}^{ab}(\tau')
	- \frac{1}{2}P(ij)P(ab)\sum_{kcd}(1 - n_b)\langle kb||cd\rangle s_i^c(\tau')s_k^a(\tau')s_j^d(\tau')\nonumber \\
	&+& \frac{1}{2}P(ij)P(ab)\sum_{klc}n_j\langle kl||cj\rangle s_i^c(\tau')s_k^a(\tau')s_l^b(\tau')
	+\frac{1}{4}P(ij)\sum_{klcd}\langle kl||cd\rangle s_i^c(\tau')s_j^d(\tau')s_{kl}^{ab}(\tau') \nonumber \\
	&+& \frac{1}{4}P(ab)\sum_{klcd}\langle kl||cd\rangle s_k^a(\tau')s_l^b(\tau')s_{ij}^{cd}(\tau') -
	P(ij)P(ab)\sum_{klcd}\langle kl||cd\rangle s_i^c(\tau')s_k^a(\tau')s_{lj}^{db}(\tau')
	\nonumber \\
	&-& P(ij)\sum_{klcd}\langle kl||cd\rangle s_k^c(\tau')s_i^d(\tau') s_{lj}^{ab}(\tau')
	- P(ab)\sum_{klcd}\langle kl||cd\rangle s_k^c(\tau')s_l^a(\tau')s_{ij}^{db}(\tau') \nonumber \\
	&+&\frac{1}{4}P(ij)P(ab)\sum_{klcd}\langle kl||cd\rangle 
	s_i^c(\tau')s_k^a(\tau')s_l^b(\tau')s_j^d(\tau').\label{eqn:S2}
\end{eqnarray}
\end{widetext}
These expressions are most easily obtained by the rules given in Section~\ref{sec:ft_cc_eq}. Note that the Fock matrix, $f$, is meant to represent only the 1st order part and therefore does not include the diagonal (orbital energies).

\section{Rules for finite-temperature, diagrammatic perturbation theory}\label{sec:Apt}
The contributions to the free energy at some finite order, $n$, in perturbation theory can be enumerated in the time domain by a diagrammatic procedure. There are many different methods for this purpose, but we will use diagrams which mimic the anti-symmetrized Goldstone diagrams common in quantum chemistry. We will imagine a time axis going from bottom to top and the basic diagrammatic components are the same as those described in Chapter 4 of Ref.~\citenum{Shavitt2009}. The $n$th order contribution to the shift in the grand potential can be obtained by the following procedure:
\begin{enumerate}
\item Draw all topologically distinct diagrams with $n$ interactions. Diagrams differing by the time-order of non-equivalent interactions are considered distinct as with other types of Goldstone diagrams.
\item Associate a unique orbital index with each directed line.
\item Associate a unique imaginary time ($\tau_1, \tau_2,\ldots$) with each interaction. 
\item With each 1-electron interaction associate a factor like $f_{pq}e^{(\varepsilon_p - \varepsilon_q)\tau}$ where $p$ is the index of the outgoing line, $q$ is the index of in-going line, and $\tau$ is the time associated with the particular interaction.
\item With each 2-electron interaction, associate a factor like $\langle pq||rs\rangle e^{(\varepsilon_p + \varepsilon_q - \varepsilon_r - \varepsilon_s)\tau}$ where $p,q,r,s$ are the indices of the left out-going, right outgoing, left incoming, and right incoming lines respectively. $\tau$ is the time associated with the interaction. 
\item Integrate each intermediate time from 0 to the next labeled time. The final time is integrated from 0 to $\beta$:
\begin{equation}
	\int_0^{\beta} d\tau_f \ldots \int_0^{\tau_3}d\tau_2\int_0^{\tau_2}d\tau_1 \ldots 
\end{equation} 
\item Sum over all orbital indices.
\item Multiply the overall diagram by a factor of $(-1)^{n-1}(-1)^{l + h}/\beta$ where $l$ is the number of closed loops and $h$ is the number of hole lines.
\item For anti-symmetrized diagrams divide by $2^s$ where $s$ is the number of pairs of equivalent fermion lines. If the standard (direct) interactions are used,  the diagram should be divided by 2 if it is symmetric with respect to reflection across a vertical line. 
\end{enumerate}
These rules can be used, at least in theory, to derive explicit expressions for the shift in the grand potential at any finite order in perturbation theory. In practice, performing the time integrals becomes increasingly cumbersome at higher order. This method can be viewed as an alternative to the frequency space method which will involve the evaluation of Matsubara sums. 

\section{The FT-CCSD $\lambda$ equations}\label{sec:Alambda}
The implementation of the FT-CCSD $\lambda$-equations mirrors that of the zero-temperature theory, but we must explicitly take into account the numerical integration scheme in order to faithfully reproduce finite difference differentiation (see Appendix~\ref{sec:Aint} for the notation and details pertaining to the numerical integration). Using a vector notation, the Lagrangian can be written as
\begin{equation}
	\mathcal{L} = \frac{1}{\beta}g_yE^y[\rv{s}^y] - 
	\frac{1}{\beta}g_y\gv{\lambda}^y\cdot \left\{\rv{s}^y +
	G_x^ye^{\gv{\Delta}(\tau^x - \tau^y)}\rv{S}^x[\rv{s}^x]\right\}
\end{equation}
where we have used the fact that all terms in the amplitude equations are evaluated at the same time. Taking the derivative with respect to a particular amplitude at a specific time point ($s_{\mu}^x$) yields an equation for the $\lambda$-amplitudes
\begin{equation}
	\lambda^{\mu x} = \frac{\partial E[\rv{s}^x]}{\partial s_{\mu}^x} -
	g_y\lambda^{\nu y} \frac{G_x^y}{g_x}e^{\Delta_{\nu}(\tau^x - \tau^y)}\frac{\partial S_{\nu}^x[\rv{s}^x]}
	{\partial s_{\mu}^x}
\end{equation}
where we have used index notation with implied summations. 
If we define a quantity
\begin{equation}
	\tilde{\lambda}^{\nu}_x \equiv g_y\lambda^{\nu y}\frac{G_x^y}{g_x}
     e^{\Delta_{\nu}(\tau^x - \tau^y)}
\end{equation}
we can write the $\lambda$ equations in a form closely resembling the zero temperature analogue:
\begin{equation}
	\lambda^{\mu x} = \frac{\partial E[\rv{s}^x]}{\partial s_{\mu}^x} -
	\tilde{\lambda}^{\nu}_x\frac{\partial S_{\nu}^x[\rv{s}^x]}
	{\partial s_{\mu}^x}.
\end{equation}

Since the amplitude equations are diagrammatically identical to the zero temperature amplitude equations, the $\lambda$ equations will also involve the same diagrams. The only difference is that we must in each iteration first compute $\tilde{\lambda}$ from $\lambda$ and then compute the new $\lambda$ amplitudes at each time point. Properties can then be evaluated by evaluating $\mathcal{L}$ with the appropriate derivative integrals. For $E$, $S$ and $N$, we require derivatives of the occupation numbers with respect to $\mu$ and $\beta$: \begin{equation}
	\frac{\partial n_p}{\partial \mu} = \beta n_p(1 - n_p) 
	\qquad\frac{\partial n_p}{\partial \beta}=(\mu - \varepsilon_p)n_p(1 - n_p).
\end{equation}
As in the zero temperature formulation, this final step can be accomplished by contraction with response-density tensors.

A slight complication arises when derivatives with respect to $\beta$ (or $T$) are required. In this case we must also consider the terms which are proportional to the derivatives of $g$ and $G$ which will in general depend on $\beta$. The specific form of these derivatives will depend on the particular quadrature scheme. In this study, we have used Simpson's rule on a uniform grid which makes these terms simple to compute. Finally, there will some contributions from the locations of the grid points which will depend on $\beta$. These contributions will vanish in the limit of a dense grid, but are necessary to faithfully reproduce the finite difference derivatives when using a small number of grid points.

\section{Numerical integration}\label{sec:Aint}
Our implementation is general enough to use a generic numerical quadrature. A function, $I(\tau)$, evaluated at the grid points will be indicated as $I_x \equiv I(\tau_x)$; the $n$ roots are labeled by $x,y,\ldots$. Integrals are then approximated as
\begin{eqnarray}
	\int_0^{\beta}I(\tau)d\tau &\approx& \sum_x g^xI_x \\
	\int_0^{\tau_y}I(\tau)d\tau &\approx& \sum_x G^x_y I_x
\end{eqnarray}
where $g$ and $G$ are the tensors of weights.

In this study we have employed a uniform grid for the sake of simplicity. For $n$ grid points, the first grid point is at $\tau = 0$, the last is at $\tau = \beta$, and the spacing between the points is given by $\delta = \beta/(n - 1)$. Simpson's rule is used for all integrations:
\begin{equation}
	\int_0^{a}I(\tau)d\tau = \frac{\delta}{3}\left[I_1 + 4I_2 + 2I_3 + 4I_4 + \ldots 
    + I_{n}\right].
\end{equation}
This defines the weights, $\rv{g}$ and $\rv{G}$. The uniform grid means we only need to perform integrals from 0 to $a$ where $a$ is a grid point, and no interpolation is required.

To compute thermodynamic quantities, we furthermore require the derivatives of the weight tensors with respect to $\beta$. Since the elements of these tensors are all linear in $\beta$, the derivative is trivial.

\bibliography{ref0,AddedRefs}
\end{document}